\documentstyle[12pt]{article}
\begin{document}

\newtheorem{theorem}{Theorem}[section]
\newtheorem{definition}[theorem]{Definition}
\newtheorem{example}[theorem]{Example}
\newtheorem{lemma}[theorem]{Lemma}
\newtheorem{proposition}[theorem]{Proposition}
\newtheorem{corollary}[theorem]{Corollary}
\newtheorem{remark}[theorem]{Remark}
\newtheorem{conjecture}[theorem]{Conjecture}

\newcommand{\A}{{\cal A}}
\newcommand{\B}{{\cal B}}
\newcommand{\al}{\alpha}
\newcommand{\Ann}{\mbox{\rm Ann}}
\newcommand{\codim}{\mbox{\rm codim}}  

\newcommand{\cn}{{{\bf {\rm C}}\hspace{-.4em}
{\vrule height1.5ex width.08em depth-.04ex}\hspace{.3em}}}
\newcommand{\Cn}{{\cn}^{n}}
\newcommand{\cpt}{\overline{\cn}^{n}}
\newcommand{\calD}{{\cal D}}
\newcommand{\Der}{\mbox{\rm Der}(s)}
\newcommand{\DG}{\cal G}
\newcommand{\hDG}{\hat{\DG}}
\newcommand{\DH}{\cal H}
\newcommand{\Dn}{\cal N}
\newcommand{\calE}{{\cal E}}
\newcommand{\End}{\mbox{\rm End\,}}
\newcommand{\gl}{{g_{\lambda}}}
\newcommand{\Gr}{\mbox{\rm Gr}}
\newcommand{\ints}{{\sf Z}\hspace{-.36em}{\sf Z}}
\newcommand{\HPA}{H_{(i,\ell), j}}
\newcommand{\HPB}{H_{(i,\ell), (j, \ell)}}
\newcommand{\HPC}{H_{(i,\ell), (j, m)}}
\newcommand{\la}{\lambda}
\newcommand{\La}{\Lambda}
\newcommand{\calL}{{\cal L}}
\newcommand{\nos}{{\rm I}\hspace{-.2em}{\rm N}}
\newcommand{\om}{\omega}
\newcommand{\oml}{\omega_{\lambda}}
\newcommand{\Om}{\Omega}
\newcommand{\Op}{\Omega^{p}}
\newcommand{\OpA}{\Omega^{p}(\A)}

\newcommand{\p}{\partial}
\newcommand{\PD}{\mbox{\rm pd\,}}
\newcommand{\calP}{{\cal P}}
\newcommand{\bfP}{{\bf P}}
\newcommand{\Pn}{{\bf P}^{n}}
 \newcommand{\Pl}{{\Phi_{\lambda}}}
\newcommand{\pl}{{\phi_{\lambda}}}
\newcommand{\Poin}{\mbox{\rm Poin}}
\newcommand{\PoinO}{\Poin(\Omega^{*}(c\A);x, }
\newcommand{\PoinGr}{\Poin(\Gr\Omega^{*}(\A); x, }
\newcommand{\proof}{{\it Proof.~}}
\newcommand{\qed}{~~\mbox{$\Box$}}
\newcommand{\R}{\rightarrow}

\newcommand{\res}{\mbox{\rm res}}
\newcommand{\ra}{{\rightarrow}}
\newcommand{\rn}{{\rm I}\hspace{-.2em}{\rm R}}

\newcommand{\scn}{\scriptsize\cn}
\newcommand{\stR}[1]{\stackrel{#1}{\longrightarrow}}
\newcommand{\th}{\theta}
\newcommand{\V}{{\cal V}}

\newcommand{\we}{\wedge}
\begin{center}
{\Large \bf Local systems over complements of hyperplanes
and}\\
\centerline{\Large \bf the Kac--Kazhdan conditions for singular vectors
}
\bigskip

\bigskip

{\sc Vadim Schechtman}
\\
{\small\it  Dept. of Mathematics, SUNY at Stony Brook, Stony Brook, NY
11794, USA}\\
{\sc  Hiroaki Terao}\\
{\small\it Dept. of Mathematics, University of Wisconsin, Madison,
WI 53706, USA}\\
{\sc  Alexander Varchenko}
\\
{\small\it Dept. of Mathematics, University of North Carolina at Chapel
Hill, NC 27599, USA}
\end{center}

In this paper we strenghten a theorem by Esnault-Schechtman-Viehweg,
\cite{ESV},
which states that one can compute the cohomology of a complement of
hyperplanes in a complex affine space with coefficients in a local system
using only logarithmic global differential forms, provided certain "Aomoto
non-resonance conditions"  for monodromies are fulfilled at some "edges"
(intersections of hyperplanes). We prove that it is enough to check
these conditions on a smaller subset of edges, see Theorem 4.1.

We show that for certain known one dimensional local systems over
configuration spaces of points in a projective line defined by a root system
and a finite set of affine weights (these local systems arise in the
geometric study of Knizhnik-Zamolodchikov differential equations,
cf. \cite{ScV}), the
Aomoto resonance conditions at non-diagonal edges coincide with Kac-Kazhdan
conditions of reducibility of Verma modules over affine Lie algebras,
see Theorem 7.1.

\section{Quasiisomorphism.}

Let $
\{H_{i} \}_{i\in I}
$ be an affine arrangement
of hyperplanes, i.e.,
$\{H_{i} \}_{i\in I}
$ is a finite collection of (distinct) hyperplanes
in the affine complex space $\cn^{n}$.
Define $U = \cn^{n} - \bigcup_{i\in I} H_{i}.$
We denote by $\Omega^{p}_{U}$ the sheaves of holomorphic
forms on $U$ for $0\leq p\leq n$.
We set ${\cal O}_{U} := \Omega^{0}_{U}.   $

For any $i\in I$, choose a degree one polynomial function
$f_{i} $ on $\cn^{n} $ whose zero locus is equal to $H_{i}. $
Define $\om_{i} := d\log f_{i} = df_{i}/f_{i} \in
\Gamma(U, \Omega^{1}_{U}). $
For a given $r\in \nos - \{0\}$ we choose matrices
$P_{i} \in \End \cn^{s}, i\in I. $
Define
\[
\om := \sum_{i\in I} \om_{i} \otimes P_{i}
\in
\Gamma (U, \Omega^{1}_{U}) \otimes \End \cn^{s}.
\]

The form $\om$ defines the connection $d + \om$ on the trivial bundle
${\cal E} := {\cal O}_{U}^{s}.  $
We suppose that $(d + \om)$ is {\it integrable} which is equivalent to
the condition $\om \wedge \om = 0$ as $d\om = 0$.
Let $\Omega^{\bullet}_{U}({\cal E}) =
\Omega^{\bullet}_{U} \otimes_{{\cal O}_{U}} {\cal E}     $
be the de Rham complex with the differential $d+\om$.

Define finite dimensional subspaces $$A^{p} \subset
\Gamma(U, \Omega^{p}_{U}({\cal E})) = \Gamma(U, \Omega^{p}_{U})
\otimes_{\bf C} \cn^{s}$$ as the $\cn$-linear subspaces generated by all
forms $\omega_{i_{1} }\wedge\cdots\wedge\omega_{i_{p} }\otimes v,
$
$v\in \cn^{s}. $
Then the exterior product by $\om$ defines
\[
A^{\bullet} : 0 \longrightarrow A^{0} \stR{\om} A^{1}
\stR{\om} \cdots \stR{\om} A^{n} \stR{} 0
\]
as a subcomplex of $\Gamma(U, \Omega^{\bullet}_{U}({\cal E}))$.

Let $
\cpt
$
be any smooth compactification of $\cn^{n} $
such that $H_{\infty}$ is a divisor.  Write
$H = H_{\infty} \cup \left(\bigcup_{\in I} H_{i} \right).$
Then $U = \cpt - H.$
(Typical examples for $\cpt$ include the complex projective space
$\bfP^{n}, $ $(\bfP^{1})^{n} $  and any toric manifold.)
Note that $\om\in \Gamma(U, \Omega^{1}_{U}) \otimes \End
\cn^{s} $ can be uniquely extended to be an $\End \cn^{s}$-
coefficient rational $1$-form $\overline{\om}$ on
$\cn^{n}$.

\begin{theorem}
\label{theorem1}
Suppose $\pi : X \rightarrow \cpt$ be a blowing up of
$\cpt$ with centers in $H$ such that 1) $X$ is nonsingular,
2) $\pi^{-1}H $ is a normal crossing divisor,
and 3) none of the eigenvalues of the resideu of
$\pi^{-1}\bar{\om} $ along any component of
$\pi^{-1}H $ lies in $\nos - \{0\}$.
Then the inclusion
\[
A^{\bullet} \hookrightarrow \Gamma (U, \Omega_{U}^{\bullet}({\cal E}))
\]
is quasiisomorphism.
\end{theorem}

\proof
Same as the proof of the first theorem in \cite{ESV}.
\qed

\section{Decomposable arrangements}

Let  $\A$ be a central
arrangement in $V$, i.e., a finite collection of
hyperplanes with $
\bigcap_{A\in \A} A
 \neq \emptyset$.
Then $\A$ is called {\bf decomposable}
if
there exist nonempty subarrangements $\A_{1} $ and $\A_{2} $
with $\A = \A_{1} \cup \A_{2} $ (disjoint) and,
after a certain linear coordinate change,
defining equations for $\A_1$ and $\A_2$ have
no common variables.

Let $\A$ be a nonempty central arrangement
in $\cn^{n} $.
Let $T =
\bigcap_{A\in \A} A
\neq \emptyset$.
Suppose $\codim T = k+1 > 0$.
Then the points of ${\bf P}_{T} := {\bf P}^{k}$
parametrize the $(\dim X + 1)$-dimensional linear
subspaces
of $\cn^{n}$ which contain $T$.
In particular, if $H$ is a hyperplane containing
$T$, it uniquely determines a
hyperplane $H'$ in
${\bf P}^{k}$.
Define $P(\A)
:=
{\bf P}^{k}  - \bigcup_{H\in\A} H'$.

\begin{definition}
\label{definition1}
Define the {\bf beta invariant} of a central arrangement
$\A$ by
\[
\beta(\A) = (-1)^{r}\chi(P(\A))
\]
where $\chi$ denotes the Euler characteristic.
\end{definition}

Let $L(\A)$ be
the set of all edges of
$\A$.  We regard $L(\A)$  as a lattice with the reverse
inclusion as its partial order.
Then $\cn^{n}$ itself is the minimum element
of $L(\A)$.  Let $\mu$ be the M\"obius function of $L(\A)$.

\begin{definition}
\label{defchar}
(\cite[Def.2.52]{OrT})
Define the characteristic polynomial of $\A$ by
\[
\chi(\A, t) = \sum_{X\in L(\A)}\mu(V, X) t^{\dim X}.
\]
\end{definition}

\begin{proposition}
\label{proposition1}
\[
\beta(\A)
=
(-1)^{k} \frac{d}{dt} \chi(\A, 1).
\]
\end{proposition}

\proof
Since $P(\A)$ is homotopy equivalent to
the complement of
of the decone
$d\A$ \cite[p.15]{OrT} of $\A$
by \cite[Prop. 2.51, Thm.5.93]{OrT}, one has
\[
(1+t) \Poin(P(\A), t)= \Poin(U, t),
\]
where $U$ is the complement of $\A$ and Poin stands for
the Poincar\'e polynomial.
Thus, by \cite[Def. 2.52]{OrT},
\begin{eqnarray*}
(t-1)^{-1} \chi(\A, t)
&=&
(t-1)^{-1} t^{\ell}\Poin(U, -t^{-1})\\
&=&
(t-1)^{-1} t^{\ell} (1-t^{-1})
\Poin(P(\A), -t^{-1}) \\
&=&
t^{\ell-1}\Poin(P(\A), -t^{-1}).
\end{eqnarray*}
Take the limit as $t$ approaches $1$.
(Note $\chi(\A, 1) = 0.$)
\qed

\medskip
Proposition \ref{proposition1} shows that
the beta invariant for the matroid
determined by $\A$.
The beta invariant for a  matroid
was introduced by
Crapo \cite{Cra}.

\begin{theorem}
\label{theorem2}
(\cite[Theorem 2]{Cra})

(1) If $\A$ is not empty, then $\beta(\A) \geq 0$.

(2) $\beta(\A) = 0$ if and only if $\A$ is decomposable.
\qed
\end{theorem}

Let $\A$ be an affine arrangement of hyperplanes in
$\cn^{n} $.
Let $L$ be an edge of $\A$.

\begin{definition}
\label{}
An edge $L$
is called {\bf dense}
in $\A$ if and only if
the central arrangement
\[
\A_{L} := \{A\in \A ~|~ L \subseteq A\}
\]
is not decomposable.
\end{definition}

By Theorem \ref{theorem2}, we have

\begin{proposition}
\label{ESVproposition}
Let $L\in L(\A)$ with
$\codim L = r + 1.$
Then the following conditions are equivalent:

(1) $L$ is dense,

(2) $\A_{L} $ is not decomposable,

(3) $\chi(P(\A_{L})) \neq 0$,

(4) $\beta(\A_{L}) := (-1)^{r} \chi(P(\A_{L})) > 0.$
\qed
\end{proposition}

\section{Resolution of a hyperplanelike divisor}
Let $Y $ be a smooth complex compact manifold
of dimension $n$, $\calD$ a divisor.
$\calD$ is {\bf hyperplanelike} if $Y$ can be covered by
coordinate charts such that in each chart $\calD$
is a union of hyperplanes.   Such charts will be called
{\bf linearizing}.

Let $\calD$ be a hyperplanelike divisor, $U$ a linearizing
chart.  A {\bf local edge} of $\calD$ in $U$ is any nonempty irreducible
intersection in $U$ of hyperplanes of $\calD$ in $U$.
An {\bf edge} of $\calD$ is the maximal analytic continuation
in $Y$ of a local edge.
Any edge is an immersed submanifold in $Y$.
An edge is called {\bf dense} if it is locally dense.

For $0\leq j\leq n-2$, let $\calL_{j} $ be the collection of
all dense edges of $\calD$ of dimension $j$.
The following theorem is essentially in
\cite[10.8]{Var}.

\begin{theorem}
\label{theorem3}
Let $W_{0} = Y$.
 Let
$\pi_1 : W_1 \rightarrow W_0$ be
the blow up along points in $\calL_0$.
In general, for $1\leq s\leq \ell-1$, let  $\pi_s : W_s  \rightarrow
W_{s-1}$ be the
blow up along the proper transforms of the $(s-1)$-dimensional
dense edges
in $\calL_{s-1}$ under $\pi_1\circ \cdots \circ \pi_{s-1}$.
Let $\pi = \pi_1\circ \cdots \circ \pi_{n-1}$.
Then $W := W_{n-1}$ is nonsingular and
$\pi^{-1} (\calD)$ normal crossing.
\end{theorem}

\section{Arrangements in ${\bf P}^{n} $ }

Let $
\{H_{i} \}_{i\in I}
$ be an affine arrangement
of hyperplanes in
 $\cn^{n}$.
Recall $U, f_{i}, \om_{i}, P_{i}, \om, \calE,$ and
$A^{\bullet} $ from Section 1.
Choose ${\bf P}^{n}$        as the compactification of
$\cn^{n}. $
Let $H_{\infty} = \Pn - \Cn$
and $\A =
\{\overline{H}_{i} \}_{i\in I}  \cup \{ H_{\infty} \}.
$
($\overline{H}_{i} $ is the closure of $H_{i} $ in $\Pn$.)
Obviously
$
\left(\bigcup_{i\in I} \overline{H}_{i} \right)  \cup H_{\infty}
$
is a hyperplanelike divisor.
Suppose $(z_{0}:\cdots :z_{n})$ be a homogeneous coordinate
system with $H_{\infty} : z_{0}=0. $
Then each $\om_{i} $ is uniquely extended to be a rational form
$\overline{\om}_{i} $ on $\Pn$;
$\overline{\om}_{i} = \om_{i} - (dz_{0}/z_{0}).$
Thus the form $\om =
\sum_{i\in I} \om_{i} \otimes P_{i}
\in
\Gamma (U, \Omega^{1}_{U}) \otimes \End \cn^{s}.
$
can be uniquely extended to $\overline{\om}$:
\[
\overline{\om}
=
\sum_{i\in I} \overline{\om}_{i} \otimes P_{i}
=
\sum_{i\in I} \om_{i} \otimes P_{i}  - (dz_{0}/z_{0}) \otimes
\left(
\sum_{i\in I}
P_{i}
\right).
\]
Define
$P_{\infty} = - \sum_{i\in I} P_{i}. $
For any edge $L$ of $\A$, let $I_{L} =
\{
i\in I\cup\{\infty\} | L\subseteq H_{i}
\}.$
Let
$P_{L} := \sum_{i\in I_{L}} P_{i}. $
By Theorems \ref{theorem1} and \ref{theorem3}, we get

\begin{theorem}
\label{theorem4}
We set $\calL$ be
the set of all dense edges of $\A$.
Suppose that

\smallskip
\noindent
{\rm {\bf (Mon)* :}}  for all $L\in \calL$, none of the eigenvalues of
$P_{L}$ lies in $\nos - \{0\}$.
\smallskip

\noindent
Then the inclusion
\[
A^{\bullet} \hookrightarrow \Gamma (U, \Omega_{U}^{\bullet}({\cal E}))
\]
is quasiisomorphism. \qed
\end{theorem}

\noindent
{\bf Remark.}  Since ``dense'' implies ``bad''
\cite{ESV}, Theorem \ref{theorem4} improves the main theorem of
\cite{ESV}.

\begin{corollary}
\label{corollary1}
Under the assumption of Theorem \ref{theorem4}, one has
\[
H^{p}(U, {\cal S}) \cong H^{p}(A^{\bullet})~~~
\mbox{\rm for~~}0\leq p\leq n
\]
where ${\cal S}$ is the local system of flat
sections of $({\cal E}, d + \omega)$
on $U$.   \qed
\end{corollary}

\begin{corollary}
\label{corollary2}
Suppose
that

\smallskip
\noindent
{\rm {\bf (Mon)** :}}  for all $L\in \calL$, none of the eigenvalues of
$P_{L} $ lies in $\nos \cup \{0\}$.
\smallskip

\noindent
Also suppose that $P_{i} P_{j} = P_{j} P_{i}$ for all $i, j$.
Then
\[
H^{p}(U, {\cal S}) = 0~~~\mbox{\rm for~~} p \neq n.
\]
\end{corollary}

\proof
By Theorem \ref{theorem4} and
\cite[4.1]{Yuz}.
\qed

\section{Discriminantal arrangements in $(\bfP^{1})^{n}$}

See \cite{ScV} for discriminantal arrangments.

Let $\Gamma$ be a graph without loops with vertices
$v_{1}, \ldots, v_{p}.  $
Let $
n_{1}$, $\ldots, n_{r}
$ be nonnegative integers,
$n =
n_{1}+\cdots + n_{r},
$
$
X = \{(i, \ell) | \ell = 1,\ldots, r, i=1,\ldots, n_{\ell} \},$
$
Y = (\bfP^{1})^{n}. $
Label the factors of $Y$ by elements of $X$
and for every $(i, \ell)\in X$ fix an affine coordinate
$t_{i}(\ell)$ on the $(i, \ell)$-th factor.

For pairwise distinct $z_{1},\ldots, z_{k} \in \cn, $
$
z_{k+1} = \infty$,
introduce in $Y$ a {\bf discriminantal arrangement}
$\A$ of ``hyperplanes''
\[
H_{(i,\ell), j}
 :
t_{i}(\ell) = z_{j}
\mbox{\rm ~for~} (i, \ell)\in X, j=1,\ldots, k+1,
\]
\[
\HPB
:
t_{i}(\ell)=t_{j}(\ell)
\mbox{\rm ~for~} 1\leq i < j \leq n_{\ell},
\]
and
\[
\HPC
:
t_{i}(\ell)=t_{j}(m)
\]
for $\ell, m$ such that $v_{\ell} $ and $v_{m} $ are joined by an edge
in the graph and
$i=1,\cdots, n_{\ell},$
$j=1,\cdots, n_{m}.$
The union of these ``hyperplanes'' is a hyperplanelike divisor.
Let $\Delta \subseteq \Gamma$ be a connected subgraph with vertices
labelled by $V \subseteq \{1,\ldots, r\}.$
For every $\ell \in V$ fix a nonempty subset
$I_{\ell} \subseteq \{1,\ldots, n_{\ell} \}.$
Fix $j\in \{1,\ldots, k+1\}$.
Introduce edges
\[
L(\{I_{\ell}\}, j)
:=
\{t\in Y ~|~ t_{i}(\ell) = z_{j} \mbox{\rm ~for~} \ell\in V, i\in
I_{\ell}  \}.
\]
Next assume that the graph $\Delta$ remains connected after any vertex
$\ell\in V$ with $|I_{\ell}| = 1$ is removed.
Under these assumptions, define edges
\[
L(\{I_{\ell}\})
:=
\{t\in Y ~|~ t_{i}(\ell) = t_{h}(\ell), t_{i}(\ell) = t_{g}(m)
\mbox{\rm ~for~} \ell, m\in V;
i, h\in
I_{\ell}; g\in I_{m}\}.
\]

\begin{proposition}
\label{proposition3}

(1)
$
L(\{I_{\ell}\}, j)
$,
$
L(\{I_{\ell}\})
$
are dense.

(2)
Every dense edge has the form above.

\end{proposition}

\proof
For any graph $G$ with vertices $\{1,\ldots, m\}$
and edges $E$, associate a central arrangement
$\A_{G} $ in $\cn^{m} $ consisting of
$\{x_{i} = 0 (1\leq i\leq m)\}$
and
$\{x_{i} = x_{j} | \{i, j\} \in E\}$.
Define a central arrangement $\B_{G} $ consisting of
$\{x_{i} = x_{j} | \{i, j\} \in E\}$.
(The arrangement $\B_{G} $ is called a graphic arrangement
\cite[2.4]{OrT}.)
In order to prove (1) and (2), it is enough to show
 the following lemma;

\begin{lemma}
\label{lemma1}

(a)
$\A_{G} $ is not decomposable iff $G$ is connected,

(b)
$\B_{G} $ is not decomposable iff $G$ is $2$--connected,
that is, $G$ remains connected after any vertex is removed.
\end{lemma}

\proof
(a):  If $G$ is disconnected, $\A_{G} $ is obviously decomposable.
If $G$ is connected, let $T$ be a maximal tree inside $G$.
Choose an edge $\{i, j\}$ such that $j$ is a terminal point of $T$.
Let $\A'$ and $\A''$ be the deletion and the restriction
of $\A_{T}$ with respect to the hyperplane $\{x_{i} = x_{j} \}.$
Since $\beta(\A') + \beta(\A'') = \beta(\A_{T})$
\cite[Theorem1]{Cra}, we can prove $\beta(\A_{T} ) = 1$
for any tree by induction on the number of edges.
This shows $\beta(\A_{G} ) \geq \beta(\A_{T} ) = 1.$

(b): Note that the matroid associated with the arrangement $\B_{G} $
is the same as the matroid associated with the graph $G$.
The matroid is connected if and only if $G$ is $2$--connected
\cite{Tut}.
\qed

Let $\Cn = Y - \bigcup_{(i, \ell)\in X} H_{(i, \ell), k+1}. $
Let $U$ be the complement in $Y$ to the union of ``hyperplanes''
of $\A$.
Recall $f_{i}, \om_{i}, P_{i}, \om, \calE,$
and $A^{\bullet} $ from Section 1.
$\om$ can be uniquely extended to be an $\End \cn^{s}$--coefficient
rational $1$-form
$
\overline{\om}
$ on $Y$.
For $(i, \ell)\in X$ the residue of
$
\overline{\om}
$ at $
H_{(i, \ell), k+1} $ is
\[
P_{(i, \ell), k+1}
=
-
\sum_{j=1}^{k}
P_{(i, \ell), j}
-
\sum_{\scriptstyle j=1 \atop
\scriptstyle j \neq i}
^{n_{\ell} }
P_{(j, \ell), (i, \ell)}
-
\sum
P_{(i, \ell), (j, m)}
\]
where the last sum is over all $m$ such that
$v_{\ell} $ and
$v_{m} $ are joined by an edge in $\Gamma$ and $j=1,\ldots,n_{m}. $

For any edge $L$ in $\A$, let $P_{L} $ be the sum of residues
of
$
\overline{\om}
$
at all ``hyperplanes'' of $\A$ contaning $L.$

\begin{theorem}
\label{theorem5}
Let $\calL$ be the set
of dense
edges of $\A$.
Suppose that

\smallskip
\noindent
{\rm {\bf (Mon)* :}}  for all $L\in \calL$, none of the eigenvalues of
$P_{L}$ lies in $\nos - \{0\}$.
\smallskip

\noindent
Then the inclusion
\[
A^{\bullet} \hookrightarrow \Gamma (U, \Omega_{U}^{\bullet}({\cal E}))
\]
is quasiisomorphism. \qed
\end{theorem}

\begin{corollary}
\label{corollary3}
Suppose
that

\smallskip
\noindent
{\rm {\bf (Mon)** :}}  for all $L\in \calL$, none of the eigenvalues of
$P_{L} $ lies in $\nos \cup \{0\}$.
\smallskip

\noindent
Also suppose that $P_{i} P_{j} = P_{j} P_{i}$ for all $i, j$.
Then
\[
H^{p}(U, {\cal S}) = 0~~~\mbox{\rm for~~} p \neq n.
\qed
\]
\end{corollary}

\section{Kac-Kazhdan conditions}
Let ${\DG}$
be a finite dimensional simple complex Lie algebra with Chevalley
generators $e_{i}, f_{i}, h_{i},\ i=1,\ldots, r.   $
Let $\DG = \Dn_{-} \oplus \DH \oplus \Dn_{+} $ be the corresponding
Cartan decomposition;
$\al_{1}, \ldots, \al_{r}\in \DH^{*}   $ the simple roots,
$\theta$ the highest root.  Let $(~,~)$ be the symmetric
non-degenerate bilinear  form on $\DG$ such that $(\theta, \theta) = 2$.

Let $T$ be an independent variable,  $\cn[T]$ the ring of polynomials,
$\cn[T,T^{-1}]$ the ring of Laurent polynomials.
For $f(T), g(T) \in\cn[T,T^{-1}]$, set
\[
\res_{0}(f(T)dg(T))  = \mbox{\rm coefficient~at~} T^{-1} \mbox{\rm ~in~}
f(T)g'(T).
\]
The space
$\DG\otimes_{\bf C} \cn[\it{T,T^{-1}}]$
is a Lie algebra with bracket
\[
[b\otimes f(T), c\otimes g(T)]
=
[b, c] \otimes f(T)g(T)
\]
for $b, c\in\DG$.
Define $\hat{\DG}$ as a central extension of
$
\DG\otimes_{\bf C} \cn[\it{T,T^{-1}}],$
\[
\hat{\DG} = \DG \otimes \cn[\it{T,T^{-1}}] \oplus \cn {\it K},
\]
where $K$ is a central element of $\hat{\DG},$
and
\[
[b\otimes f(T), c\otimes g(T)]
=
[b, c]\otimes f(T)g(T)
+
(b, c) \res_{0} (f(T)dg(T))K.
\]
Set $\hDG^{+} =\DG \otimes \cn[\it T]\oplus \cn {\it K}$; it is a Lie
subalgebra of $\hDG$.

Fix a complex number $k$. Set $\kappa=k+g$ where $g$ is the dual Coxeter
number of $\DG$, cf. \cite{K}, 6.1.

For $\Lambda \in
\DH^{*}, $ let $M(\La)$ be the Verma module over $\DG$
with highest weight $\La$.
Consider $M(\La)$ as a $\hDG^{+}$-module by setting
$
\DG \otimes \it{T}\cn[\it T]
$
to act as zero and $K$ as multiplication by $k$.
Set
\[
\hat{M}(\La) := U(\hDG) \otimes_{U(\hDG^{+})} M(\La).
\]
It is a Verma module over $\hDG$.

\begin{proposition}
\label{proposition4}
(Kac-Kazhdan conditions)
$\hat{M}(\La)
$
is reducible if and only if
at least one of the following
three conditions is satisfied.

(1) $\kappa = 0$.

(2) There exist a positive root $\al$ of $\DG$
and natural numbers $p, s \in \nos - \{0\}$
such that
\[
(\Lambda, \al) + (\rho, \al)
=
p \frac{(\al, \al)}{2} - (s-1) \kappa,
\]
where $\rho$ is half-sum of positive roots
of $\DG$.

(3) There exist a positive root $\al$ of $\DG$
and natural numbers $p, s \in \nos - \{0\}$
such that
\[
(\Lambda, \al) + (\rho, \al)
=
- p \frac{(\al, \al)}{2} + s \kappa.
\]
\end{proposition}

\proof
We use notations of \cite{K}, Ch. 6,7. In these notations the Kac-Kazhdan
reducibility condition, \cite{KK}, Thm 1, reads as
$$
\left<\La,\nu^{-1}(\beta)\right>+ \left<\hat{\rho},\nu^{-1}(\beta)\right>
-
p\frac{(\beta,\beta)}{2}=0
$$
for some positive root $\beta$ of $\hDG$ and a positive integer $p$.
(Here we denoted by $\hat{\rho}$ an element denoted by $\rho$ in \cite{K},
to distinguish it from our $\rho$.)

By {\em loc. cit.}, 6.3, every
such $\beta$ has one of the following forms:
(1) $\beta=m\delta,\ m>0$; (2) $\beta=\alpha+m\delta,\ m\geq 0$;
(3) $\beta=-\alpha +m\delta,\ m>0$, where $\alpha$ is a positive root
of $\DG$, $m$ an integer. From {\em loc. cit} it follows easily that
$\left<\Lambda,\nu^{-1}(\delta)\right>
=k$, $\left<\hat{\rho},\nu^{-1}(\delta)\right>=g$ and
$\left<\hat{\rho},\nu^{-1}(\alpha)\right>=(\rho,\alpha)$.
This implies the proposition.
\qed
\medskip

Let $w$ be the longest element of the
Weyl group of $\DG$.
For $\La\in\DH^{*}, $
set $\La' = -w(\La)$.

\begin{proposition}
\label{proposition5}
$\hat{M}(\La')$ is reducible if and only if
$\hat{M}(\La) $
is reducible.  The Kac-Kazhdan conditions
for $\La'$ expressed in terms of $\La$
coincide with the Kac-Kazhdan conditions
for $\La$.
\end{proposition}

\proof
For a positive root $\al$, $-w(\al)$ is a positive root.
This implies the proposition.
\qed

\section{Resonances of discriminantal arrangements}

Let $\Gamma$ be the Dynkin diagram of a complex simple
Lie algebra $\DG.$
The vertices of the diagram are labelled by simple roots
$\al_{1},\ldots, \al_{r}  $ of the algebra.  Let $n_{1},\ldots, n_{r}  $
be nonnegative integers, $n = n_{1} +\cdots + n_{r}. $
For pairwise distinct $z_{1},\ldots, z_{k}\in \cn, z_{k+1}=\infty, $
consider in $Y = (\bfP^{1})^{n} $ the discriminantal arrangement
$\A$ associated to these data.

Let $\La_{1},\ldots, \La_{k} \in \DH^{*}. $
Set
$
\La_{k+1} =
-\om (\La_{1} +\cdots +\La_{k} - n_{1} \al_{1} -\cdots - n_{r}\al_{r}).  $
Fix a nonzero complex number $\kappa$.
Introduce an integrable connection $d + \om$ on the trivial bundle
$\calE := {\cal O}_{U} $ with
\[
\om = \sum_{(i,\ell)\in X}\sum_{j=1}^{k}
P_{(i, \ell), j} \om_{(i, \ell), j}
+  \sum_{\ell=1}^{r} \sum_{1 \leq i < j \leq n_{\ell}}
P_{(i, \ell), (j, \ell)} \om_{(i, \ell), (j, \ell)}
 +  \sum_{1\leq \ell < m \leq r} \sum_{i=1}^{n_{\ell}}
\sum_{j=1}^{n_{m}}
P_{(i, \ell), (j, m)} \om_{(i, \ell), (j, m)},
\]
where
\[
\om_{(i, \ell), j}
=
d(t_{i}(\ell) - z_{j})
/
(t_{i}(\ell) - z_{j}),
{}~~
\om_{(i, \ell), (j, m)}
=
d(t_{i}(\ell) - t_{j}(m))
/
(t_{i}(\ell) - t_{j}(m)),
\]
\[
P_{(i, \ell), j} = -(\al_{\ell}, \La_{j})/\kappa,
{}~~
P_{(i, \ell), (j, m)} = -(\al_{\ell}, \al_{m})/\kappa,
\]
see \cite{ScV} and \cite{Var}.
$\om$ extends to be a rational $1$-form $\overline{\om}$ on $Y$.

For any edge $L$ in $\A$, let $P_{L} $ be the sum of residues of
$\overline{\om}$ at all ``hyperplanes''
of $\A$ containing $L$.
For $p\in\nos \cup \{0\}$, we say that the connection $d+\om$
has a {\bf resonance at} $L$ {\bf of level} $p$, if $P_{L} = p.$

The following theorem connects resonances of $\A$ with the Kac-Kazhdan
conditions for the Verma modules
$
\hat{M}(\La_{1}),
\ldots,
\hat{M}(\La_{k+1})
$
of the affine algebra $\hDG.$
Let $\al = \sum a_{\ell} \al_{\ell} $
be a positive root of $\DG,$ $p$ a natural number.
Assume that $a_{\ell} p \leq n_{\ell} $ for all $\ell$.
For every $\ell$, fix a subset $I_{\ell} \subseteq
\{1,\ldots, n_{\ell} \}$ consisting of $a_{\ell} p$
elements.

\begin{theorem}
\label{theorem6}

(1) For every $j = 1,\ldots, k+1,$ the edge $L_{j}
= L(\{I_{\ell}\}, j)$ is dense.

(2) For $j = 1,\ldots, k$ and every natural number $s$,
the resonance condition at $L_{j} $ of level $ps$,
$P_{L_{j} } = ps,$ coincides with the Kac-Kazhdan
condition of type (2) for $\hat{M}(\La_{j}),$
\[
(\La_{j}, \al) + (\rho, \al)
=
p \frac{(\al, \al)}{2} - s\kappa.
\]
(3) For $j = k+1$ and every natural number $s$,
the resonance condition at $L_{k+1} $ of level $ps$,
$P_{L_{k+1} } = ps,$ coincides with the Kac-Kazhdan
condition of type (3) for $\hat{M}(\La_{k+1}),$
\[
(\La_{k+1}, \al) + (\rho, \al)
=
- p \frac{(\al, \al)}{2} + s\kappa.
\]
\end{theorem}

{\bf Remarks.}
(1) For resonance values of $\La_{1},\ldots ,\La_{k}, \kappa,$
nontrivial cohomological relations occur in the image of
$A^{\bullet} \subset \Gamma(U, \Omega_{U}(\calE))$.
The Theorem suggests that the relations correspond to singular
vectors in the Verma modules $\hat{M}(\La_{1}), \ldots,
\hat{M}(\La_{k+1})$.
In \cite{FSV}
this correspondence was established for the simplest singular
vector in $\hat{M}(\La_{k+1}),$
the correspondence implied algebraic equations
satisfied by conformal blocks in the WZW model of conformal
field theory.

(2) For $j = 1,\ldots, k$ and natural number $p$, the Kac-Kazhdan
condition,
$
(\La_{j}, \al)
+
(\rho, \al)
=
p
\frac{(\al, \al)}{2},
$
appears as a degeneration condition for a certain contravariant
form of the arrangement $\A$, see \cite[Secs. 3, 6]{ScV}.

\medskip

\proof
(1) For a positive root $\al = \sum a_{\ell} \al_{\ell} $
consider the subset $\{\al_{\ell}~|~a_{\ell} > 0\}$ of the set
of simple roots.  The subset distinguishes a subgraph of the Dynkin
diagram.  The subgraph is connected \cite[ch. 7, sec. 1]{Bou}.
Now $L_{j} $ is dense by Proposition \ref{proposition3}.

(2)
\begin{eqnarray*}
P_{L_{j}} - ps
&=& \frac{1}{\kappa}
[
(-\La_{j}, \al) p + \sum_{r=1}^{r}
\frac{p a_{\ell} (p a_{\ell} - 1) }{2} (\al_{\ell}, \al_{\ell})
 +
\sum_{1\leq \ell < m \leq r} p a_{\ell} p a_{m} (\al_{\ell}, \al_{m})
]
- ps\\
&=&
\frac{p}{\kappa}
\left[
-(\La_{j}, \al) + p \frac{(\al, \al)}{2} -
\sum_{\ell=1}^{r} a_{\ell} \frac{(\al_{\ell}, \al_{\ell})}{2}
-s \kappa
\right]\\
&=&
\frac{p}{\kappa}
\left[
-(\La_{j}, \al) - (\rho, \al) + p
\frac{(\al_{\ell}, \al_{\ell})}{2}
-s \kappa
\right].
\end{eqnarray*}
This proves (2).  Part (3) is proved by similar direct computations
using Proposition \ref{proposition5}.
\qed

\medskip
{\small \bf Acknowledgement.}
{\small The authors thank H\'el\`ene Esnault
for pointing out an error in an earlier version.
The second author thanks G. Ziegler for directing
his attention to the work of H. Crapo about the beta invariant.
He is also thankful to A. Libgober and S. Yuzvinsky for useful
discussions.}

\end{document}